\documentclass[dvips]{acta}
\usepackage{supertabular,lscape,epsfig}

\usepackage{amssymb}
\usepackage{polski}
\usepackage{amsmath}

\SetPages{1}{11}

\SetVol{61}{2011}

\begin{document}

\def\degr{\hbox{$^\circ$}}
\def\arcmin{\hbox{$^\prime$}}
\def\arcsec{\hbox{$^{\prime\prime}$}}
\def\fd{\hbox{$.\!\!^{\rm d}$}}
\def\fh{\hbox{$.\!\!^{\rm h}$}}
\def\fm{\hbox{$.\!\!^{\rm m}$}}
\def\fs{\hbox{$.\!\!^{\rm s}$}}
\def\fdg{\hbox{$.\!\!^\circ$}}
\def\farcm{\hbox{$.\mkern-4mu^\prime$}}
\def\farcs{\hbox{$.\!\!^{\prime\prime}$}}

\begin{Titlepage}
\Title{J1145-0033 $-$ the most distant giant radio source?}

\Author{A. Ku\'zmicz, E. Kuligowska, M. Jamrozy}
{Astronomical Observatory, Jagiellonian University, ul. Orla 171, 30-244, Krak{\'o}w, Poland\\
e-mail:(cygnus, elzbieta, jamrozy)@oa.uj.edu.pl}

\Received{March 07, 2011}
\end{Titlepage}

\Abstract{We present J1145-0033, a candidate for the most distant (z=2.055) lobe-dominated giant radio quasar, with a projected linear size of 1.34 Mpc. This quasar has both FR II -- type radio morphology and broad absorption lines in its optical spectrum. Some physical characteristics (e.g. black hole mass, accretion rate, equipartition magnetic field, energy density and particle density of ambient medium) based on the optical and radio data are provided. We have also found that the quasar has a relatively large central black hole mass and a very small accretion rate in comparison with similar objects.}{Galaxies: active -- quasars: evolution -- intergalactic medium}

\section{Introduction}
 
\noindent
The observed giant radio sources (GRSs; for references see e.g. Ischwara-Chandra \& Saikia 1999) with linear size $\gtrsim$1~Mpc are powerful extragalactic radio-emitting objects at the final stages of their life and activity, which makes them a valuable tool for studying evolution of active galactic nuclei (AGNs). For several years GRSs were not supposed to be found at redshifts higher than z$\sim$1, because of the expected strong density increase of  intergalactic medium (IGM). Kapahi (1989) showed that the IGM density evolves as $\rho_{\rm IGM}\propto (1+z)^{3}$. Therefore, the large environmental densities hamper the radio structure linear-size evolution at high redshift. However, Law-Green et al. (1995) discovered a GRS (4C39.24) hosted by a galaxy located at z=1.883. Moreover, a sample of relatively distant (0.3$<$z$<$0.9) and large radio galaxies were presented by Cotter et al. (1996). It seems that GRSs could be used to probe the cosmological evolution of IGM even up to redshifts as high as z$>$0.5 (Machalski et al. 2007). The disadvantage of using GRSs as cosmological density determinants is the small number of known sources of such a type. It is mostly because distant GRSs are not easy to be identified at the modern interferometric radio survey maps available. Detecting steep-spectrum and low surface-brightness radio-bridges connecting the radio core with hot spots for distant GRSs is a quite challenging task but it would be possibly facilitated with the advent of novel low-frequency telescopes such as the Low Frequency Array (LOFAR) and the Square Kilometre Array (SKA). There are a number of efforts under way aiming to increase the number of high-z GRSs and a sample of the largest radio sources (predominantly quasars; QSOs) with 1$<$z$<$2 was presented by Kuligowska et al. (2009).\\ 
In this paper we present the source J1145-0033, a recently discovered candidate for the most distant GRS. It has a 1.34 Mpc{\footnote{Throughout this paper we assumed the $\rm \Lambda$CDM cosmology with $\rm\Omega_{m}$ = 0.27, $\rm\Omega_{\Lambda}$= 0.73 and H$\rm_{0}$ = 71 km s$^{-1}$ Mpc$^{-1}$.}} structure with a typical  FRII (Fanaroff \& Riley, 1974) morphology and is hosted by a QSO. This giant radio QSO (GRQ) was identified in the process of systematic search for high-redshift (z$>$1) GRSs (for details see Kuligowska et al. 2009) on the basis of the optical spectra  taken from the Sloan Digital Sky Survey (SDSS; Adelman-McCarthy et al. 2007) and the radio maps from the Faint Images of the Radio Sky at Twenty-Centimeters survey (FIRST; Becker et al. 1995).

\section{J1145-0033: radio data}

\noindent
J1145-0033 was mapped in the 1.4-GHz FIRST VLA survey. It has a compact radio core of 3.86 mJy at RA: 11$\rm ^{h}$45$\rm ^{m}$53\fs69, Dec: -00$\rm ^{o}$33$^\prime$4\farcs84 (J2000.0). The possible hot-spots of 6.77 and 3.94 mJy are located at RA: 11$\rm ^{h}$45$\rm ^{m}$49\fs62, Dec: -00$\rm ^{o}$33$^\prime$36\farcs14 
and RA: 11$\rm ^{h}$45$\rm ^{m}$58\fs40, Dec: -00$\rm ^{o}$32$^\prime$10\farcs32, respectively. The south-western hot-spot is located 1\farcm15 away from the core and the arm-length ratio of the lobes is 1.29. J1145-0033 is also visible as a 14.52 mJy triple source in the 1.4-GHz NRAO VLA Sky Survey (NVSS; Condon et al. 1998). Figure 1 presents the contour maps from the NVSS and FIRST surveys overlaid on the optical SDSS image. The host QSO of J1145-0033 is labeled as `A' in this figure. The maps as well as the measured values of basic radio parameters (e.g. the projected linear size D, the core and total radio fluxes S) were processed using the AIPS package. The NVSS contours display a weak structure around the core, oriented at an angle of about 70$\rm ^{o}$ to the symmetry axis of the main radio structure. This could be a remnant of some backflow. Using the NVSS maps of Q and U Stokes, we obtained a radio polarimetric map, which shows a strong polarization of the south-western lobe with a polarized flux of 1.93 mJy. This gives a fractional polarization ratio of about 24.8\%. The electric field vectors are all oriented in almost the same direction, forming an angle of about 30$\rm ^{o}$ with the main symmetry axis of the source. No polarized flux from the north-eastern lobe and the central region have been detected; however, it could be just a purely instrumental effect due to beam depolarization.  

Using the 1.4 GHz fluxes we derived the corresponding total (P$\rm _t$) and core (P$\rm _c$) radio power values using the formula given by Brown et al. (2001):\\
\\
$\rm log P(WHz^{-1})=log S(mJy)-(1+\alpha) \cdot log(1+z)+2log(D_L[Mpc])+17.08$\\
\\
where $\alpha$ is the spectral index ($S\sim\nu^{\alpha}$) of the source and $\rm D_L$ denotes its luminosity distance. Since we had only one-frequency radio data of our source available, we adopted spectral index values  of $\alpha = -0.6$ and $\alpha = -0.3$ for the lobes and the core, respectively, provided by Wardle et al. (1997) and Zhang \& Fan (2003) as typical mean values of radio QSOs. We estimated also the inclination angle i of the radio structure in respect to the line--of--sight using the formula:
\begin{center}
$\rm i=[acos(\frac{1}{\beta_j} \cdot \frac{(s-1)}{(s+1)})]$ \\
\end{center}
where $\rm s=(S_{j}/S_{cj})^{1/2-\alpha}$, $\rm S_{j}$ is the flux of the lobe, $\rm S_{cj}$ is the flux of the counter-lobe, and $\rm \beta_j$ is the jet velocity. Following Wardle et al. (1997), we assumed $\rm \beta_j=0.6$c. The obtained values of radio power and inclination angle are provided in Table 1.\\

\begin{figure}[t]
\begin{center}
\includegraphics[width=12cm, angle=-90]{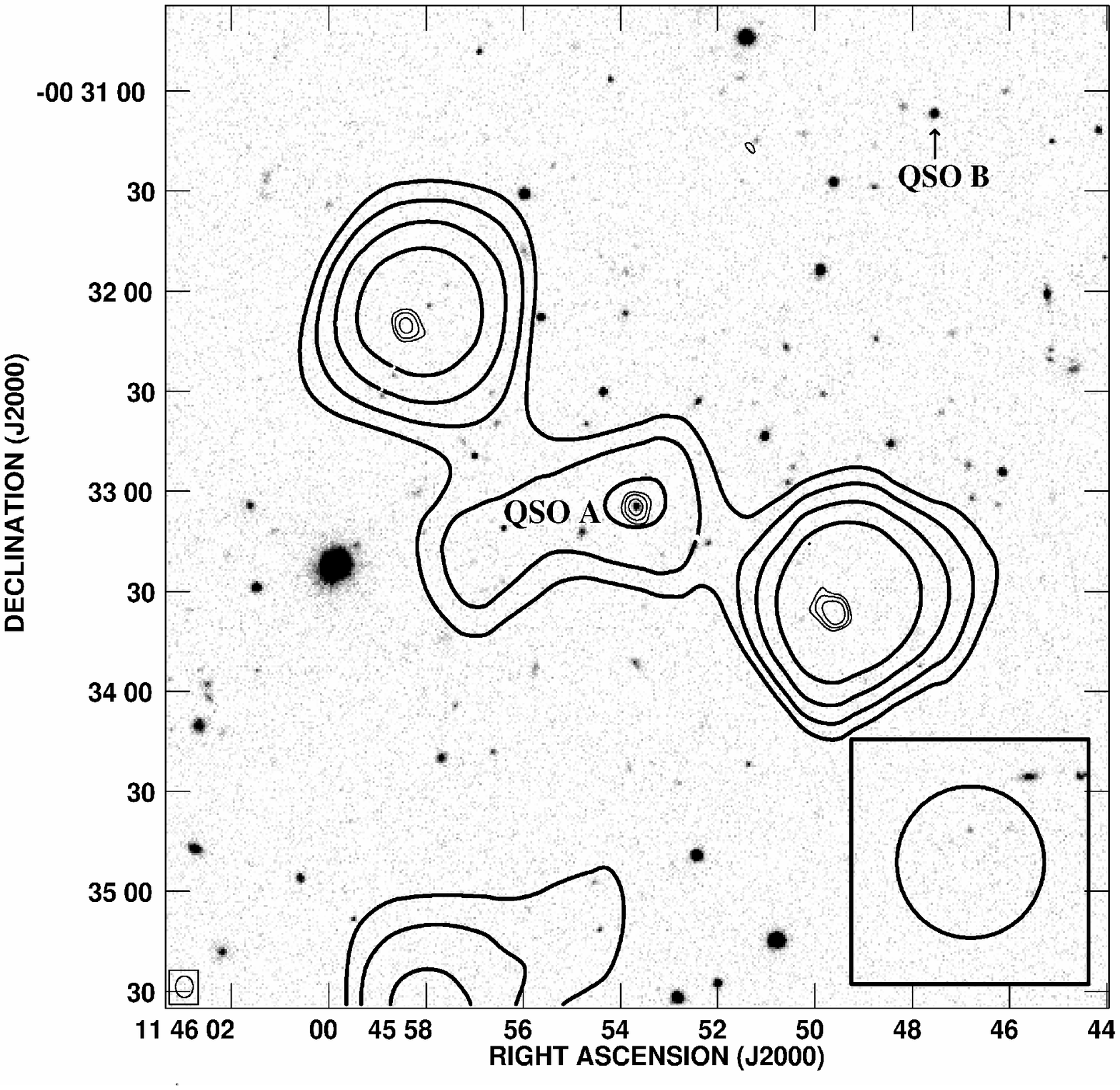}
\FigCap{The 1.4-GHz VLA maps of the GRQ J1145-0033 taken from the NVSS survey (bold contours) and from the FIRST survey (thin contours) overlaid on the optical image from the SDSS. The contour levels are: 1, 1,41, 2, 2,83$\times$1 mJy/beam and 1, 2, 4$\times$0.6 mJy/beam for the NVSS and FIRST surveys, respectively. The ellipses in the right and left bottom corners represent the resolution of the FIRST and NVSS surveys, respectively. Labels QSO `A' and QSO `B' mark the position of the J1145-0033 QSO and the companion J1145-0031 QSO, respectively.}
\end{center}
\end{figure}

\section{J1145-0033: optical data}

\noindent
The parent object of J1145-0033 is a QSO located at redshift z=2.05446$\pm$0.00270 (according to the data from the 7$\rm {th}$ SDSS release). Its coordinates are given in Table 1. The difference between the radio and the optical position is 
0\farcs26 and 0\farcs24, in right ascension and declination, respectively. The observed SDSS optical magnitudes of the QSO are: u=20.55$\pm$0.06, g=20.01$\pm$0.02, r=19.63$\pm$0.02, i=19.26$\pm$0.02, and z=18.97$\pm$0.04. Its optical spectrum  taken from the SDSS was reduced through the standard procedures (galactic extinction correction, redshift correction) with the IRAF package. Next we fitted a power-law continuum in the spectral windows where there were no emission lines observed (i.e. 1320 -1350\AA, 1430-1460\AA, and 1790-1830\AA) and then used the iron template in the UV band taken from Vestergaard \& Wilkes (2001) to fit the Fe continuum. The fitting procedure was similar to that described by Boroson \& Green (1992). The result of continuum subtraction is presented in Figure 2. 

After the continuum subtraction, we measured FWHM of the CIV (1549 \AA) line. We did not fit any line profile, because the CIV line in QSOs spectra almost always has an asymmetric profile, which is usually neither Gaussian nor Lorentzian. The highly ionized CIV line shows a systematic blueshift (Richards et al. 2002) due to the absorption (occultation) of the CIV red wing. Therefore, we measured the line FWHM in the same way as Peterson et al. (2004). The corresponding black hole (BH) mass was computed using the mass-scaling relation taken from Vestergaard \& Peterson (2006): 
\begin{center}
$\rm M_{BH}(CIV1549) = 4.57 \cdot 10^6 (\frac{\lambda L_{\lambda}(1350\AA)}{10^{44}erg s^{-1}})^{0.53\pm0.06} \cdot (\frac{FWHM(CIV1549)}{1000 km s^{-1}})^2 M_{\odot}$,\\
\end{center}
where ($\rm \lambda L_{\lambda}$) is the monochromatic continuum luminosity at 1350 \AA. Using the BH mass estimations, it is easy do determine the accretion rate (\.m) onto the BH. The accretion rate is defined as \.m = $\rm L_{bol}$/$\rm L_{Edd}$ where $\rm L_{bol}$ is the QSO bolometric luminosity:\\

$\rm L_{bol} = 4.6 \rm \times$ $\rm \lambda L_{\lambda}$(1350\AA); (Vestergaard 2004)\\
\\
and $\rm L_{Edd}$ is Eddington luminosity:\\

$\rm L_{Edd}=1.45\cdot10^{38}M_{BH}/M_{\odot}ergs^{-1}$; (Dietrich et al. 2009).\\
\\ 
The obtained BH mass for our QSO of $\rm 10^{9.27\pm0.6}M_{\odot}$ is relatively high in comparison to BH masses 
of other QSOs (e.g. from Wilhite et al. 2007). Its Eddington luminosity is $\rm 10^{47.43}$ erg/s and the accretion rate is 0.013. We compared also the above results with similar ones obtained for a sample of radio QSOs (Ku\'zmicz \& Jamrozy, in preparation) and discussed them in Sect.~4.5. In Figure 3 we present relations between redshift, BH masses, and accretion rate.

\begin{figure}[t]
\begin{center}
\includegraphics[width=13cm]{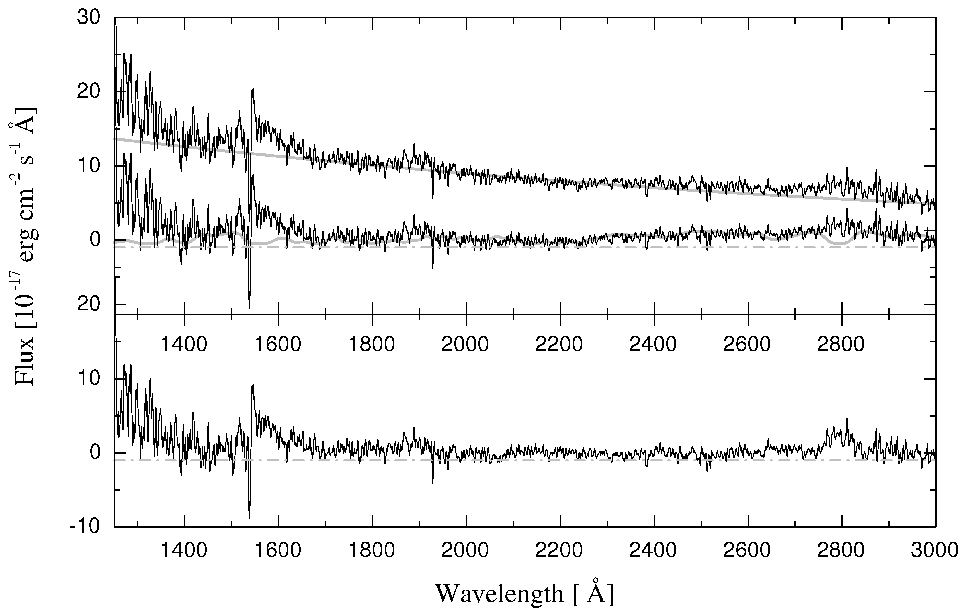}
\FigCap{The optical spectrum of the host QSO of J1145-0033. {\bf upper panel:} top spectrum is the observed spectrum in the rest frame with a power-law continuum overlaid. The bottom spectrum is the continuum-subtracted spectrum and the best fit of the Fe emission. {\bf bottom panel:} the spectrum of the QSO after power-law and Fe continuum subtraction.}
\end{center}
\end{figure}

\begin{figure}[t]
\begin{center}
\includegraphics[width=10cm]{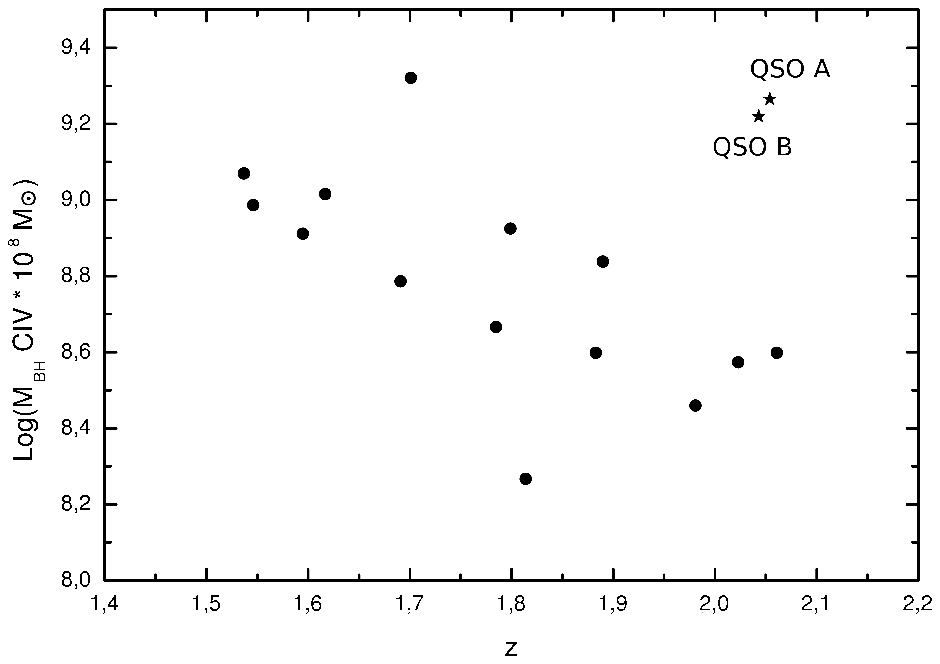}\\
\includegraphics[width=10cm]{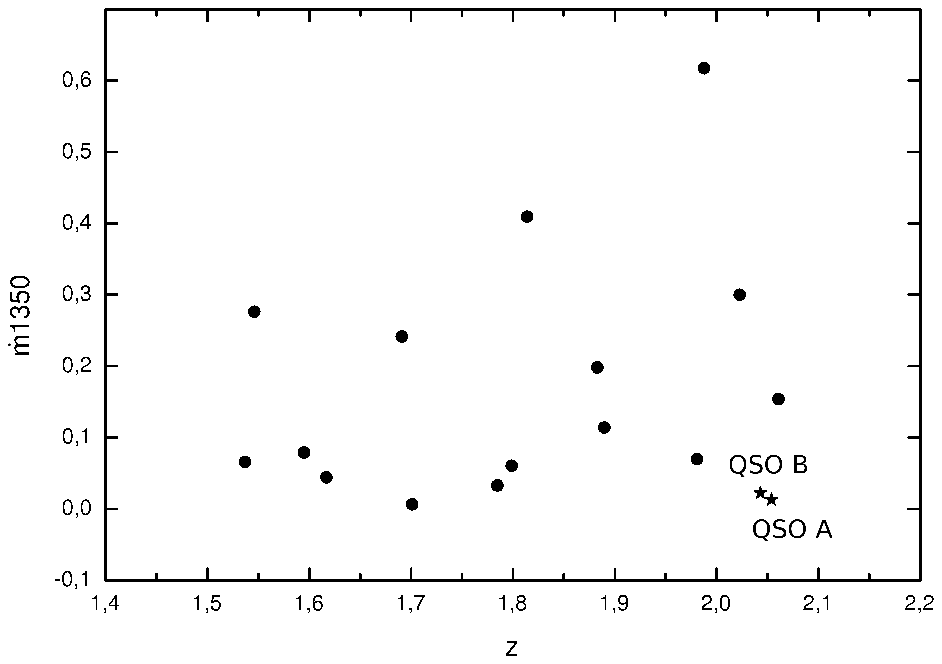}
\FigCap{Black hole masses and accretion rates for radio QSOs. {\bf upper panel:} relation between the BH mass and redshift.{\bf bottom panel:} accretion rate as a function of redshift. QSO A and B are marked as asterisks and the other radio QSOs as filled circles.}
\end{center}
\end{figure}

\vspace{-0.5cm}
\MakeTable{lc}{5.5cm}{The observational and derived parameters of GRQ J1145$-$0033 and the companion QSO J1145-0031.}
{\hline
Parameter                            &\hspace{1cm}Value    \\
\hline
\bf{QSO A}&\\
IAU name                             &\hspace{1cm} J1145$-$0033\\
$\alpha$ (J2000) [h m s]             &\hspace{1cm} 11 45 53.67\\
$\delta$ (J2000) [$\rm ^o$ $'$ $''$] &\hspace{1cm} $-$00 33 04.6\\
z                                    &\hspace{1cm} 2.055\\
d [arcmin]                           &\hspace{1cm} 2.642\\
D [Mpc]                              &\hspace{1cm} 1.340 \\
$\rm S_{1.4 GHz }$ total (NVSS) [mJy] &\hspace{1cm} 14.52\\
$\rm S_{1.4 GHz }$ core (FIRST) [mJy] &\hspace{1cm} 3.86\\
Log $\rm P_{1.4 GHz}$ total [W/Hz]   &\hspace{1cm} 26.47\\
Log $\rm P_{1.4 GHz}$core [W/Hz]     &\hspace{1cm} 25.72\\
Log$\rm M_{BH}(CIV) [10^{8} M_{\odot}]$ &\hspace{1cm} 9.27$\pm0.60$\\
i [$\rm^o$]                          &\hspace{1cm} 82\\
FWHM$\rm (CIV)$ [\AA]               &\hspace{1cm} 61.12\\
Log $\rm \lambda L_\lambda$ (1350\AA)&\hspace{1cm} 44.867\\
\.m                                  &\hspace{1cm} 0.013\\
BI [km/s]                            &\hspace{1cm} 0\\
AI [km/s]                            &\hspace{1cm} 1576\\
\hline
\bf{QSO B}&\\
IAU name                             &\hspace{1cm} J1145$-$0031\\
$\alpha$ (J2000) [h m s]             &\hspace{1cm} 11 45 47.55\\
$\delta$ (J2000) [$\rm ^o$ $'$ $''$] &\hspace{1cm} $-$00 31 06.7\\
z                                    &\hspace{1cm} 2.043\\
Log$\rm M_{BH}(CIV) [10^{8} M_{\odot}]$ &\hspace{1cm} 9.22$\pm0.04$\\
FWHM$\rm (CIV)$ [\AA]                &\hspace{1cm} 51.4\\
Log $\rm \lambda L_\lambda$ (1350\AA)&\hspace{1cm} 45.06\\
\.m                                  &\hspace{1cm} 0.022\\
\hline   
\hline
}

\section{Discussion}

\subsection{Companion QSO}

\noindent
The host QSO of J1145-0033 (thereafter named as QSO `A') has a `companion' QSO, first recognized by Kirkman \& Tytler (2008) who studied the absorption in a large sample of QSO pairs (separated in the plane of sky $<$3 Mpc). The companion QSO J1145-0031 (thereafter named as QSO `B') is located at RA: 11$\rm ^{h}$45$\rm ^{m}$47\fs55, Dec: -00$\rm ^{o}$31$^\prime$6\farcs72 (J2000.0). It has a redshift of 2.04295$\pm$0.00320 (according to the data from the 7$\rm ^{th}$ SDSS release) and optical magnitudes u=18.89 $\pm$0.02, g=18.70 $\pm$0.01, r=18.70 $\pm$0.01, i=18.57 $\pm$0.01,  and z=18.34 $\pm$0.02. Using similar methods as for the QSO `A', we estimated the mass of the companion QSO `B' to be $\rm 10^{9.22\pm0.04}M_{\odot}$. Wilhite et al. (2007) found the mass to be $\rm 10^{9.26\pm0.02}M_{\odot}$, which is quite consistent with our result. Only 29 QSOs out of 615 from the Wilhite et al. (2007) sample have  BH masses larger than that of the QSO `B', and the maximal BH mass they estimated is $\rm 10^{9.55}M_{\odot}$. The Eddington luminosity of the QSO `B' is $\rm 10^{47.39}$ erg/s and its accretion rate is 0.022. Contrary to the GRQ, the QSO `B' is a radio-quiet object.\\
The separation between the J1145-0033 and J1145-0031 QSOs is 149$^{\prime\prime}$, which gives the actual distance between them as 1.23 Mpc. This separation is only $\sim$1.58 times larger than the distance from the Milky Way to the M31 galaxy (0,784 Mpc; Stanek \& Garnavich 1998) . While the separation of our binary QSO is too large for direct gravitational interactions to trigger accretion, it still could be located in an overdense environment, where interactions with other galaxies can support its activity (e.g., Djorgovski 1991,  Hopkins et al. 2008). The idea of using QSOs to indicate dense regions in the early universe is not new (e.g. Djorgovski 1999, Djorgovski et al. 1999). While it is unclear to date whether high-redshift QSOs reside in protoclusters, compelling evidence for overdensities has been found around a handful of z $>$ 4 radio galaxies (e.g., Hennawi et al. 2010). Absorption measurements in the vicinity of QSO pairs can be used for determining IGM density in the corresponding regions (Kirkman \& Tytler 2008, Hennawi et al. 2006). The absorption seems to be much higher than that calculated solely from the QSO luminosity (Guimaraes et al. 2007). This result implies that the QSOs are situated in regions where the IGM is overdense by the factor of $\sim$5 (Guimaraes et al. 2007). There are also other hints (e.g. correlation between small-scale excess of galaxy and QSO clustering) suggesting that the QSOs are likely to be found in dense environments (Bowen et al. 2006, Hennawi et al. 2006). If this is the case also in the vicinity of our QSO pair, it may be hard to explain how the radio structure of J1145-0033 evolved to a Mpc scale in overdense environment. We searched the SDSS database looking for more companion objects of similar redshift as the QSOs `A' and `B' within the circle of 19\farcm7 (which corresponds to 10 Mpc) in diameter around QSOs pair, but no such objects have been found.   

\subsection{Broad absorption line QSOs}

\noindent
Interestingly enough, the optical spectrum of J1145-0033 shows high ionization broad absorption lines (HiBAL), which makes it a member of a rare category of QSOs (Trump et al. 2006). The BAL classification is usually based on a value of the balnicity index (BI; Weymann 1991) that defines the strength of its absorption features. In the case of QSO `A', the classification was performed using an absorption index (AI; Trump et al. 2006), which is more sensitive on narrower absorption lines. For our QSO the BI=0, but AI=1576km/s (Trump et al. 2006).\\

The BAL phenomenon is observed in about 10$-$20\% (depending on the selection criteria) of the entire QSO population (Weymann 2002; Tolea et al. 2002, Hewett \& Foltz 2003, Reichard et al. 2003, Trump et al. 2006). BAL are probably caused by high-velocity gas outflows during the accretion processes. There have been proposed two scenarios to account for this phenomenon. In the first scenario, BAL regions exist in both BAL and non-BAL QSOs, and the BAL QSOs are just normal QSOs but seen along a particular line-of-sight (Weymann et al. 1991). The second scenario states that BAL are present only during a relative short (possibly episodic) evolutionary phase of QSO activity, which occurs most likely at an early stage of their evolution (e.g. Becker et al. 2000). Only a small fraction among BAL QSOs are radio-loud objects and most of them have core-dominated radio morphologies. They belong to the compact steep spectrum (CSS) and gigahertz-peaked spectrum (GPS) objects, which are considered to be young radio sources of linear sizes less than 20 kpc (Kunert-Bajraszewska et al. 2010). The BAL QSOs with extended radio structures are very rare. Only eight BAL QSOs with extended FRII radio structure are known to date (Greeg et al. 2006). In almost all cases their projected linear size is within the range of 117$-$585 kpc, and one of them (J1408+3054, Greeg et al. 2006) is considered to be a GRQ with the linear size of 1.65 Mpc.

As it was shown in Sect. 2, the inclination angle, i, of the J1145-0033 radio structure is about 82$\rm ^o$, which means that the lobes lie almost in the celestial plane. This can suggest that  BAL could be just due to orientation of the source. Such an orientation implies that the gas outflows could be accounted for by a radiatively accelerated wind from the accretion disc or gas evaporating from a dusty torus (Punsly 2006). While this explanation seems quite plausible, the rarity of FR II BAL QSOs and their observed anticorrelation between the BI and the radio loudness (Gregg et al. 2006) may not allow for its confirmation. Brotherton et al. (2006) found that the polar outflows are present  also in FR II radio QSOs (e.g. PKS 0040-005), thus both the facts can support the alternative origin of BAL, related with relatively short (episodic) evolutionary phase (e.g. Gregg et al. 2006). This scenario is also consistent with the result obtained for QSO pairs by Kirkman \& Tytler (2008), who found that the QSOs display episodic activity with timescales of 0.3 -- 10 Myr.

\subsection{The radio structure}

\noindent
As it was mentioned in the Introduction, the strong cosmological size evolution of powerful radio sources and the decrease in surface brightness with redshift ($\rm \propto (1+z)^{-4}$) make the detection of extended sources difficult. 
Moreover, the suppression of bridge emission by inverse Compton losses against the cosmic microwave background (CMB) increases strongly with redshift. Therefore, the `tailless hot-spots' of large sources at high redshifts could be easily mistaken for unrelated sources. The magnetic field equivalent to the microwave background at the redshift of J1145-0033 is $\sim2.8$nT, $\sim 5$ larger than the minimum magnetic field of this source (for details of the calculation see Sect. 4.4). Certainly the energy losses in this GRQ will be dominated by the inverse Compton scattering. Emission from inverse Compton scattered CMB photons, in the form of diffuse, extended X-rays between the radio core and hot-spots, has been detected in a number of GRSs at large redshift (z$\gtrsim$1; Erlund et al 2008; Laskar et al 2010). The three components of J1145-0033 can indeed appear as isolated sources. However, the major axis of the western hot-spot is aligned with the compact core. If the lobe-components were actually foreground unrelated sources, we would expect them to have visible optical counterparts. We cross-checked the radio positions of the hot-spots with the SDSS image, but no such objects up to the limiting SDSS magnitude  could be identified. The alignment of two side sources with the core, symmetry of their radio flux, as well as the small value ($\sim1$) of arm-length ratio of the both lobes, provide evidence that the three components are elements of one structure. However, some doubts are raised  by the large asymmetry of polarized emission between the two lobes. In order to confirm the radio structure,  it is necessary to perform deep multifrequency radio observations, particularly at low frequencies.

\subsection{The IGM}

\noindent
The pressure of adiabatically expanding non-relativistic uniform IGM at a particular redshift can be derived using the equation $\rm p_{IGM}(z)=p_{IGM}(z=0)(1+z)^{5}$ with the the present value of IGM density 
$\rm p_{IGM}(z=0)\sim 10^{-14}dyn\,cm^{-2}$ (Subrahmanyan \& Saripalli 1993; Schoenmakers et al. 2000). Assuming that the bridges of radio galaxies are in static pressure equilibrium with the ambient IGM, it is possible to estimate the minimum internal pressure of any particular radio source. In the location of J1145-0033, the pressure of ambient IGM should be 
$\rm\sim3\times10^{-12}dyn\,cm^{-2}$. On the other hand, the minimum energy density within a radio source is directly related to its pressure, $\rm p_{min}=(\gamma -1)\times u_{min}$, were $\gamma$ is the ratio of specific heats. In the case of ultrarelativistic gas, $\gamma=\frac{4}{3}$, and so $\rm p_{min}=\frac{1}{3}u_{min}$. In order to estimate the energy density and the magnetic field one usually has to rely on minimum energy or equipartition arguments. We calculated these parameters following Longair (1997). The parameters entering the estimation have been taken as: the cutoff frequencies $\nu_{min}=10$ MHz, $\nu_{max}=100$ GHz, the filling factor $\phi$=1, ratio of the energy content in relativistic protons to that of electrons $\eta$=1, and the spectral index $\alpha$=$-$0.6. The source volume has been calculated assuming a cylindrical shape for the radio lobes and a rude estimation gives V$\rm=4.8\times10^{70}$ $\rm cm^{3}$. For the GRQ we obtained $\rm u_{min}=0.31\times10^{-11}dyn\,cm^{-2}$. The pressure estimated assuming minimum energy conditions is comparable (differ only by a factor of three) with the pressure estimated for uniform intergalactic medium and its $\sim(1+z)^{5}$ cosmological evolution. The particle densities of ambient medium is $\rm n[cm^{-3}] = u_{min}/3kT$. Here k is the Boltzmann constant and T the temperature of the IGM (in K). For temperatures of $10^{7}$ K, it gives $\rm n = 7.4\times 10^{-4}cm^{-3}$.

\subsection{Black hole masses and accretion rates} 

\noindent
The obtained values for the BH masses of QSO `A' and `B' are relatively high and their accretion rates are small in 
comparison with other BHs from a sample of QSOs with extended radio emission but smaller linear size ($<$ 1 Mpc) 
(Ku\'zmicz \& Jamrozy, in prep.). It can be clearly seen in Fig.~3 that the characteristics of these two QSOs differ from the general trends for the other sources in the $\rm M_{BH}$-z and \.m $-$z planes. The parameter values of both the QSOs are, however, very similar to each other, which may suggest that they are distinguished among other QSOs 
with regard to either some internal properties of their host galaxies or external properties of their common IGM. Their evolution could be much similar, except for the presence of radio emission from one of them. The small accretion rates of 0.013 
and 0.022, respectively for QSOs `A' and `B', are close to the lower limit of radio-loud QSOs (which is $\sim $0.01; Gu et al. 2001). The large BH mass and small accretion rate of these QSOs could be explained, according to Netzer et al. (2007), through occurring of at least one earlier episode of faster BH growth with a high ($\sim 1$) accretion rate. There are several examples of radio galaxies to show structures of multi-episodes of AGN activity (Saikia \& Jamrozy 2009) and  there is one QSO 4C02.27 (Jamrozy et al. 2009) to display radio structures originating from two different cycles of nuclear activity among them.  

The problem of BH mass grow was investigated by a number of authors by studying large samples of optically selected QSOs.
Usually, as a result they obtained BH mass increasing with redshift to an upper limit of $\sim10 \rm ^{10} M_{\odot}$ 
(at z$\sim$ 2) and then leveling out (e.g. Shen et al. 2008). Such a trend could be strongly biased by selection effects arising from flux-limited samples. From our analysis performed for a sample of lobe-dominated QSOs, we have obtained a quite opposite relation, where the BH mass decreases with redshift. This could be much more easily accounted for in the frame of the $\Lambda$CDM cosmological model, with small structures forming originally and then evolving to larger ones by the way of consecutive mergers.

\section{Summary}

\noindent
We have presented here J1145-0033, a candidate for the most distant GRS, which additionally displays properties of a HiBAL QSO, as well as possesses a companion QSO 1.23 Mpc away. The obtained values for the BH masses of both these QSOs are relatively high and their accretion rates are small in comparison with other BHs from a sample of QSOs with extended radio emission. In order to confirm the radio structure and to determine other physical parameters of this source, e.g. its dynamical and spectral age, jet power, or environmental density, it is however necessary to perform additional deep multifrequency radio observations. 

\vspace{0.5cm}
{\bf ACKNOWLEDGMENTS}\\
We are grateful to Marianne Vestergaard for providing us with the template of Fe emission. 
We thank Staszek Zo{\l}a for helpful comments on the manuscript.
This project was funded by the MNiSW grant 3812/B/H03/2009/36.

\end{document}